\documentclass[twocolumn,prl,showpacs,floatfix]{revtex4}

\usepackage{graphicx}
\usepackage{dcolumn}
\usepackage{bm}
\usepackage{amsmath}

\begin{document}

\title{Trial wave functions, molecular states, and ro-vibrational spectra in 
the\\ lowest Landau level: A universal description for bosons and fermions}

\author{Constantine Yannouleas and Uzi Landman}

\affiliation{School of Physics, Georgia Institute of Technology,
             Atlanta, Georgia 30332-0430}


\begin{abstract}
Through the introduction of a class of appropriate translationally invariant
trial wave functions, we show that the strong correlations in the lowest Landau 
level (LLL) reflect in finite systems the emergence of intrinsic point-group 
symmetries associated with rotations and vibrations of molecules formed through 
particle localization. This molecular description is universal, being valid for 
both bosons and fermions, for both the yrast and excited states of the LLL 
spectra, and for both low and high angular momenta. This physical picture is
fundamentally different from the "quantum-fluid" one associated with 
Jastrow-type trial functions. 
\end{abstract}


\maketitle

{\it Motivation.\/} $-$ Following the discovery \cite{tsui82} of the fractional 
quantum Hall effect (FQHE) in two-dimensional (2D) semiconductor 
heterostructures under high magnetic fields ($B$), the description of 
strongly correlated electrons in the lowest Landau level (LLL) developed 
into a major branch of theoretical condensed matter physics \cite{laug8399,
hald83,halp84,trug85,jain89,moor91,ston92,oakn95,pala96,jainbook,
yann02,yann0304,yann07,chan05,chan06}.
Early on, it was realized that the essential many-body physics in the LLL 
could be captured through trial wave functions. Prominent examples are 
the Jastrow-type Laughlin \cite{laug8399}, composite fermion \cite{jain89}, 
and Moore and Read's \cite{moor91} Pfaffian functions, representing 
quantum-liquid states \cite{laug8399}. More recently, the field of 
semiconductor quantum dots \cite{yann07} helped to focus attention on finite 
systems with a small number ($N$) of electrons. Theoretical investigations of
such finite systems led to the introduction of "crystalline"-type LLL trial 
functions referred to as rotating electron molecules (REMs) 
\cite{yann02,yann07}. 

Most significantly, the burgeoning field of trapped ultracold 
neutral gases has generated recently an unparalleled interest regarding the 
fundamental aspects (including the appropriateness of trial 
functions) of strongly correlated states in the lowest Landau level 
\cite{mott99,gunn00,gunn01,pape01,ueda01,popp04,joli05,barb06,baks07,coop08}; 
in this case, the LLL manifold of degenerate single-particle orbitals arises 
as a result of the rapid rotation (with rotational frequency $\Omega$) of the 
trap. Furthermore, it is anticipated that small assemblies of ultracold bosonic 
atoms will become technically available in the near future \cite{geme08} and 
that they will provide an excellent vehicle \cite{popp04,barb06,coop08,geme08} 
for experimentally reaching exotic physical behavior beyond the mean-field 
Gross-Pitaevskii regime and for testing the rich variety of proposed LLL trial 
wave functions. 

A universal description of the full LLL spectra (including both yrast 
\cite{note3} and all excited states), however, is still missing. To remedy this,
a unified theory for the LLL spectra of a small number of particles valid for 
both statistics (i.e., for both bosons and fermions) is introduced in this 
paper. The LLL spectra are shown to be associated with {\it fully quantal\/} 
ro-vibrational molecular (RVM) states, i.e., states 
described by trial wave functions akin to the REM functions of Ref.\ 
\cite{yann02}. It is remarkable that the numerical results of the present 
theory agree within machine precision with exact-diagonalization (EXD) 
results, including energies, wave functions, and overlaps. This behavior
outperforms the behavior of all other trial functions, including those of the
composite-fermion view \cite{chan05,chan06,jeon07}. 
The RVM functions and corresponding point-group symmetries express the emergent 
many-body intrinsic structure of the highly correlated LLL states.

{\it Theory.\/} $-$ The RVM functions have the general form
(within a normalization constant):
\begin{equation}
\Phi^{\text{RXM}}_{\cal L}(n_1,n_2) Q_\lambda^m |0>, 
\label{mol_trial_wf}
\end{equation}
where $(n_1,n_2)$ indicates the molecular configuration (here we consider 
two concentric rings) of point-like particles with $n_1$ ($n_2$) 
particles in the first (second) ring. The particles on each ring form 
regular polygons. The index RXM stands 
for either REM, i.e., a rotating electron molecule, or RBM, i.e., a rotating 
boson molecule. $\Phi^{\text{RXM}}_{\cal L}(n_1,n_2)$ 
alone describes pure molecular rotations associated with magic angular momenta 
${\cal L}={\cal L}_0+n_1k_1 + n_2k_2$, with $k_1$, $k_2$ being nonnegative 
integers; ${\cal L}_0=N(N-1)/2$ for electrons and ${\cal L}_0=0$ for bosons. 
The product in Eq.\ (\ref{mol_trial_wf}) combines rotations 
with vibrational excitations, the latter being 
denoted by $Q_\lambda^m$, with $\lambda$ being an angular momentum; the 
superscript denotes raising to a power $m$. Both $\Phi^{\text{RXM}}_{\cal L}$ 
and $Q_\lambda^m$ are homogeneous polynomials of the complex particle coordinates 
$z_1,z_2,\ldots,z_N$, of order ${\cal L}$ and $\lambda m$, respectively.
The total angular momentum $L={\cal L}+\lambda m$. $Q_\lambda^m$ is always 
symmetric in these variables; $\Phi^{\text{RXM}}_{\cal L}$ is antisymmetric 
(symmetric) for fermions (bosons). $|0> = \prod_{i=1}^N \exp[-z_iz_i^*/2]$; this
product of Gaussians will be omitted henceforth.

The analytic expressions for the $\Phi^{\text{REM}}_{\cal L}$ (for fully 
polarized electrons) were derived in Ref.\ \cite{yann02} employing a 
two-step method: (i) First a single Slater {\it determinant\/} [that breaks the 
rotational (circular) symmetry] was constructed using displaced Gaussians as 
electronic orbitals, i.e.,  
\begin{eqnarray}
u(z,Z_j) = \frac{1}{\sqrt{\pi}} \exp[-|z-Z_j|^2/2] \exp[-i (xY_j+yX_j)].
\label{gaus}
\end{eqnarray}
The phase factor is due to the gauge invariance. $z \equiv x-i y$,
and all lengths are in dimensionless units of ${l_B}\sqrt{2}$, with 
the magnetic length $l_B=\sqrt{\hbar /(m_e \omega_c)}$;  
$\omega_c = eB/(m_ec)$ is the cyclotron frequency. The centers 
$Z_j \equiv X_j+i Y_j$, $j=1,2,\ldots,N$ of the Gaussians are the vertices of the 
regular polygons in the $(n_1,n_2)$ geometric arrangement. (ii) A subsequent step 
of symmetry restoration was performed using the projection operator
${\cal P}({\cal L})=\frac{1}{2\pi} \int_0^{2 \pi} d\gamma 
e^{i\gamma(\hat{L}-{\cal L})}$, where $\hat{L}=\sum_{i=1}^N \hat{l}_i$ is the 
total angular momentum operator; this yielded trial wave functions with good total 
angular momenta ${\cal L}$ \cite{yann02,yann07}.

Analytic expressions for the $\Phi^{\text{RBM}}_{\cal L}$ (for spinless bosons)
can also be derived using the two-step method. Naturally, in the first step
one constructs a {\it permanent} out of the orbitals of Eq.\ (\ref{gaus}); one also
uses the equivalence $\omega_c \rightarrow 2\Omega$ between the 
cyclotron frequency (electrons) and the rotational frequency (bosons) 
\cite{yann07}. The expressions for any number $N$ of bosons and any
molecular configuration $(n_1,n_2, \ldots ,n_q)$ will be pressented in Ref.\ 
\cite{yann09}. Here we present as an illustrative example the simpler case of $N=3$
bosons having a $(0,3)$ one-ring molecular configuration. One has (within a 
normalization constant)  
\begin{equation}
\Phi^{\text{RBM}}_{\cal L}(0,3) = 
\sum_{0 \leq l_1 \leq l_2 \leq l_3}^{l_1+l_2+l_3={\cal L}}
C(l_1,l_2,l_3) \;{\text{Perm}} [z_1^{l_1}, z_2^{l_2}, z_3^{l_3}],
\label{rbm1}
\end{equation}
where the symbol "Perm" denotes a permanent with elements $z_i^{l_j}$, $i,j=1,2,3$; 
only the diagonal elements are shown in Eq.\ (\ref{rbm1}). The coefficients
were found to be:
\begin{eqnarray}
C(l_1,l_2,l_3) &=& \left(\prod_{i=1}^3 l_i! \right)^{-1} 
\left(\prod_{k=1}^M p_k! \right)^{-1}
\nonumber \\
&\times& \left( \sum_{1 \leq i < j \leq 3} 
\cos \left[\frac{2\pi(l_i-l_j)}{3} \right] \right),
\label{rbm2}
\end{eqnarray} 
where $1 \leq M \leq 3$ denotes the number of different
indices in the triad $(l_1,l_2,l_3)$ and the $p_k$'s are the 
multiplicities of each one of the different indices. For example, 
for (1,1,4), one has $M=2$ and $p_1=2$, $p_2=1$. 

The $\Phi^{\text{REM}}_{\cal L}$ expressions for electrons in a $(0,N)$ 
or a $(1,N-1)$ configuration are given by Eqs.\ (2) and (4) of Ref.\ 
\cite{yann0304}, respectively. For electrons (1) $M=N$ in all instances and (2) a 
{\it product of sine\/} terms replaces the {\it sum of cosine\/} terms appearing in
Eq.\ (\ref{rbm2}). 

We note that $\Phi^{\text{RXM}}_{\cal L}(n_1,n_2) =0$ for both bosons and 
electrons when ${\cal L} \neq {\cal L}_0 + n_1 k_1 + n_2 k_2$. This selection rule 
follows directly from the point group symmetries of the $(n_1,n_2)$ molecular
configurations.

The vibrational excitations $Q_\lambda$ are given by the same expression for
both bosons and electrons, namely, by the symmetric polynomials:
\begin{equation}
Q_\lambda = \sum_{i=1}^N (z_i-z_c)^\lambda,
\label{ql} 
\end{equation}
where $z_c=(1/N)\sum_{i=1}^N z_i$ is the coordinate of the center of mass and 
$\lambda>1$ is a prime number. Vibrational excitations of a similar form, i.e., 
$\tilde{Q}_\lambda=\sum_{i=1}^N z_i^\lambda$ (and certain other variations), 
have been used earlier to approximate {\it part\/} of the LLL spectra. Such 
earlier endeavors provided valuable insights, but overall they remained 
inconclusive; for electrons over the maximum density droplet [with magic 
${\cal L}={\cal L}_0$], see Refs.\ \cite{ston92} and \cite{oakn95}; for electrons 
over the $\nu=1/3$ ($\nu={\cal L}_0/{\cal L}$) Jastrow-Laughlin trial function 
[with magic ${\cal L}=3 {\cal L}_0$], see Ref.\ \cite{pala96}; and for bosons in 
the range $0 \leq L \leq N$, see Refs.\ \cite{mott99,ueda01,pape01}. 

The advantage of expression (\ref{ql}) \cite{note1} is that it is translationally 
invariant (TI), a property also shared by both the $\Phi^{\text{RBM}}_{\cal L}$ and 
$\Phi^{\text{REM}}_{\cal L}$ trial functions. 
In the following, we will discuss illustrative cases, which will demonstrate that
the molecular trial functions of Eq.\ (\ref{mol_trial_wf}) provide a correlated
basis that spans the TI subspace   
\cite{trug85,pape01,vief08} of {\it nonspurious\/} states in the LLL spectra. 
The dimension $D^{\text{TI}}(L)$ of the TI subspace is much
smaller than the dimension $D^{\text{EXD}}(L)$ of the exact-diagonalization (EXD) 
space spanned by the uncorrelated determinants ${\text{Det}} 
[z_1^{l_1}, \ldots, z_N^{l_N}]$ or permanents ${\text{Perm}} [z_1^{l_1}, \ldots, 
z_N^{l_N}]$. The remaining $D^{\text{EXD}}(L) - D^{\text{TI}}(L)$ states are 
{\it spurious\/} center-of-mass excitations (generated by applying 
$\tilde{Q}_1^m$) whose energies coincide with those 
appearing at all the other smaller angular momenta \cite{trug85}. Thus 
$D^{\text{TI}}(L)=D^{\text{EXD}}(L)- D^{\text{EXD}}(L-1)$; see Tables 
\ref{ene_bos_np3} and \ref{ene_ferm_np4}. 
 


{\it Three spinless bosons\/} $-$ Only the $(0,3)$ molecular 
configuration and the dipolar $\lambda=2$ vibrations are at play
(as checked numerically), i.e.,
the full TI spectra at any $L$ are spanned by the wave functions 
\begin{equation}
\Phi^{\text{RBM}}_{3k}(0,3) Q_2^m \Rightarrow \{k,m\},
\label{wfbos3}
\end{equation} 
with $k,m=0,1,2,\ldots$, and $L=3k+2m$; these states 
are always orthogonal. This represents a remarkable analogy with the case of $N=3$ 
electrons (see below). 

\begin{table*}[t] 
\caption{\label{ene_bos_np3}%
Spectra of three spinless bosons. Second column: Dimensions of the EXD and the
nonspurious TI (in parenthesis) spaces. Fourth to sixth columns: Matrix elements 
[in units of $g/(\pi \Lambda^2)$, $\Lambda=\sqrt{\hbar/(m\Omega)}$] of the 
repulsive contact interaction $g \delta(z_i-z_j)$ 
between the nonspurious states $\{k,m \}$ [see Eq.\
(\ref{wfbos3})]. The total angular momentum $L=3k+2m$. Last three columns: Total 
energy eigenvalues from the diagonalization of the associated matrix of 
dimension $D^{\text{TI}}(L)$. There is no nonspurious state with $L=1$. The 
full spectrum at a given $L$ is constructed by including, in addition to the
listed TI total-energy eigenvalues [$D^{\text{TI}}(L)$ in number], all the
energies associated with angular momenta smaller than $L$.
An integer in square brackets indicates the energy ordering
in the full spectrum, with [1] denoting an yrast state. 
Seven decimal digits are displayed, but the total energies agree with the 
EXD ones within machine precision.}
\begin{ruledtabular}
\begin{tabular}{rllllllll}
$L$ & $D^{\text{EXD}}(D^{\text{TI}})$ & $\{k,m\}$ & 
\multicolumn{3}{l}{Matrix elements} & 
\multicolumn{3}{l}{Total energy eigenvalues (TI)} \\ \hline
0 & 1(1)  & \{0,0\} & 1.5000000  & 
  ~~~~      &  ~~~~ & 1.5000000[1] & ~~~~ & ~~~~\\
2 & 2(1) & \{0,1\} & 0.7500000  &
  ~~~~      & ~~~~ & 0.7500000[1] & ~~~~ & ~~~~\\
3 & 3(1) & \{1,0\} & 0.3750000  &
  ~~~~      & ~~~~ & 0.3750000[1] & ~~~~ & ~~~~\\
4 & 4(1) & \{0,2\} & 0.5625000  &
  ~~~~      & ~~~~ & 0.5625000[2] & ~~~~ & ~~~~\\
5 & 5(1) & \{1,1\} & 0.4687500  & 
  ~~~~      & ~~~~ & 0.4687500[2] & ~~~~ & ~~~~\\
6 & 7(2) & \{2,0\} & 0.0468750  & 
 0.1482318  &  ~~~~ &  ~~~~       & ~~~~ & ~~~~\\
~~~   & ~ & \{0,3\} & 0.1482318  & 0.4687500   & ~~~~ & 0.0000000[1] & 
0.5156250[4] & ~~~~\\
7 & 8(1) & \{1,2\} & 0.4921875  & 
  ~~~~      & ~~~~ & 0.4921875[4] & ~~~~ & ~~~~\\
8 & 10(2) & \{2,1\} & 0.0937500  & 
0.1960922   &~~~~ &  ~~~~         & ~~~~ & ~~~~\\
~~~   & ~ & \{0,4\} & 0.1960922  & 0.4101562  & ~~~~ & 0.0000000 & 
0.5039062[6] & ~~~~\\
12 & 19(3) & \{4,0\} & 7.3242187$\times 10^{-4}$ & 1.0863572$\times 10^{-2}$ & 
1.5742811$\times 10^{-2}$ & ~~~~~~  & ~~~~~~  & ~~~~~ \\
~~~ & ~~~ & \{2,3\} & 1.0863572$\times 10^{-2}$ & 0.1611328 & 
0.2335036 & ~~~~~~ & ~~~~~  & ~~~~~ \\
~~~ & ~~~ & \{0,6\} & 1.5742811$\times 10^{-2}$ & 0.2335036 & 
0.3383789 & 0.0000000  & 0.0000000  & 0.5002441[13] \\
\end{tabular}
\end{ruledtabular}
\end{table*}

Table \ref{ene_bos_np3} provides the systematics of the molecular
description for the beginning ($0 \leq L \leq 12$) of the LLL spectrum. 
There are several cases when the TI
subspace has dimension one and the exact solution coincides with a single
$\{k,m\}$ state. For $L=0$ the exact solution coincides with 
$\Phi^{\text{RBM}}_0=1$ ($Q_\lambda^0=1$); 
this is the only case when an LLL state has a Gross-Pitaevskii form, i.e., it is
a single permanent [see $|0\rangle$ in Eq.\ (\ref{mol_trial_wf})].
For $L=2$, we found $\Phi^{\text{exact}}_{[1]} \propto Q_2$ (for the subscript 
$[i]$, see caption of Table \ref{ene_bos_np3}), and since [see Eq.\ (\ref{ql})]
$Q_2 \propto(z_1-z_c)(z_2-z_c)+(z_1-z_c)(z_3-z_c)+(z_2-z_c)(z_3-z_c)$, this 
result agrees with the findings of Refs.\ \cite{smit00,pape01} concerning 
ground states of bosons in the range $0 \leq L \leq N$. For $L=3$, one finds
$\Phi^{\text{exact}}_{[1]} \propto \Phi^{\text{RBM}}_3$. Since 
$\Phi^{\text{RBM}}_3 \propto(z_1-z_c)(z_2-z_c)(z_3-z_c)$ [see Eq. (\ref{rbm1})], 
this result agrees again with the findings of Refs.\ \cite{smit00,pape01}. 
For $L=5$, the single nonspurious state is an excited one, 
$\Phi^{\text{exact}}_{[2]} \propto \Phi^{\text{RBM}}_3 Q_2$.
For $L=6$, the ground-state is found to be $\Phi^{\text{exact}}_{[1]} \propto
-160\Phi^{\text{RBM}}_6/9  + Q_2^3/4=(z_1-z_2)^2(z_1-z_3)^2(z_2-z_3)^2$, i.e., 
the bosonic Laughlin function for $\nu=1/2$ is equivalent to an RBM state that 
incorporates vibrational correlations. For $L \geq N(N-1)$ (i.e.,
$\nu \leq 1/2$), the EXD yrast 
energies equal zero, and with increasing $L$ the degeneracy of the zero-energy 
states for a given $L$ increases. It is important that this nontrivial behavior 
\cite{wenbook} is reproduced faithfully by the present method (see Table 
\ref{ene_bos_np3}).

{\it Three electrons\/} $-$ Although unrecognized, the solution of the
problem of three spin-polarized electrons in the LLL using molecular trial 
functions has been presented in Ref.\ \cite{laug83.2}. 
Indeed, the wave functions in Jacobi 
coordinates in Eq. (18) of Ref.\ \cite{laug83.2} are precisely of the form 
$\Phi^{\text{REM}}_{3k} Q_2^m$, as can be checked after transforming back to 
cartesian coordinates \cite{note2}. 
It is noteworthy that Laughlin did not present 
molecular trial functions for electrons with $N >3$, or for bosons for any $N$. 
This is done in the present paper. We further note that the 
well-known Jastrow-Laughlin 
ansatz $\prod_{1\leq i<j \leq N} (z_i-z_j)^{2p+1}$ introduced in Ref.\ 
\cite{laug8399} is described as a quantum-fluid state \cite{laug8399}. The 
quantum-liquid-picture interpretation of this ansatz is inconsistent with 
the RVM functions of Eq.\ (\ref{mol_trial_wf}).

{\it Four electrons\/} $-$ For $N=4$ spin-polarized electrons, one needs to
consider rovibrational states [see Eq.\ (\ref{mol_trial_wf})] for two distinct 
molecular configurations, i.e., $\Phi^{\text{REM}}_{6+4k}(0,4) Q_\lambda^m$ and 
$\Phi^{\text{REM}}_{6+3k}(1,3) Q_\lambda^m$. 
Vibrational excitations with $\lambda \geq 2$ must also be
considered. In this case the molecular basis states are not always orthogonal,
and the Gram-Schmidt orthogonalization is implemented. Table \ref{ene_ferm_np4} 
summarizes the molecular description in the start of the LLL spectrum
($6 \leq L \leq 15$ and $L=18$). 
 
\begin{table*}[t] 
\caption{\label{ene_ferm_np4}%
Spectra of four spin-polarized electrons. 
Second column: Dimensions of the EXD and the nonspurious TI (in parenthesis) spaces.
Last three columns: Total energy 
eigenvalues [in units of $e^2/(\kappa l_B)$] from the diagonalization of the 
Coulomb interaction $e^2/(\kappa r_{ij})$ in the TI subspace spanned by the trial 
functions $\Phi^{\text{REM}}_{6+4k}(0,4) 
Q_\lambda^m$ and $\Phi^{\text{REM}}_{6+3k}(1,3) Q_\lambda^m$. 
Third to sixth columns: the molecular configurations $(n_1,n_2)$ and the quantum 
numbers $k$, $\lambda$ and $m$ are indicated within brackets. 
There is no nonspurious state with 
$L=7$. An integer in square brackets indicates the energy ordering of each 
nonspurious state in the full spectrum, obtained by considering the spurious 
center-of-mass excitations (see also caption of Table \ref{ene_bos_np3}).
Eight decimal digits are displayed, but the total energies agree 
with the corresponding EXD results within machine precision.}
\begin{ruledtabular}
\begin{tabular}{llllll}
$L$ & $D^{\text{EXD}}$($D^{\text{TI}}$)& $[(n_1,n_2)\{ k, \lambda, m \}]$ &
\multicolumn{3}{l}{Total energy eigenvalues (TI)} \\
\hline
6 & 1(1) & [(0,4)\{0,$\lambda$,0\}] & 
2.22725097[1]  &   ~~~~      & ~~~~~ \\
8 & 2(1) & [(0,4)\{0,2,1\}] & 
2.09240211[1] &   ~~~~      & ~~~~~~ \\
9 & 3(1) & [(1,3)\{1,$\lambda$,0\}] &
 1.93480798[1] &   ~~~~      & ~~~~ \\
10 & 5(2) & [(0,4)\{1,$\lambda$,0\}] [(0,4)\{0,2,2\}] & 
 1.78508849[1]  &  1.97809256[3] & ~~~~ \\
11 & 6(1) & [(1,3)\{1,2,1\}] & 1.86157215[2]  & 
  ~~~~      & ~~~~~ \\
12 & 9(3) &  [(0,4)\{1,2,1\}] [(0,4)\{0,2,3\}] [(1,3)\{2,$\lambda$,0\}] &
 1.68518201[1] & 1.76757420[2]  & 1.88068652[5] \\
13 & 11(2) & [(1,3)\{1,2,2\}] [(0,4)\{1,3,1\}]  & 1.64156849[1]  & 
 1.79962234[5] & ~~~~ \\
14 & 15(4) & [(0,4)\{2,$\lambda$,0\}] [(0,4)\{1,2,2\}] [(0,4)\{0,2,4\}] & 
 1.50065835[1]  & 1.63572496[2]   &  1.72910626[5]  \\
~~~ & ~~~~ & [(1,3)\{2,2,1\}] & 
 1.79894008[8] & ~~~~~   & ~~~~~~ \\
15 & 18(3) & [(1,3)\{3,$\lambda$,0\}] [(1,3)\{2,3,1\}] [(1,3)\{1,3,2\}]  & 
 1.52704695[2] & 1.62342533[3] & 1.74810279[8] \\
18 & 34(7) & [(0,4)\{3,$\lambda$,0\}] [(0,4)\{2,2,2\}] [(0,4)\{1,2,4\}] & 
 1.30572905[1]  &  1.41507954[2]   &  1.43427543[4] \\
~~~~ & ~~~~ & [(0,4)\{0,2,6\}] [(1,3)\{4,$\lambda$,0\}] [(1,3)\{2,2,3\}]  & 
 1.50366728[8]  &   1.56527615[11]   & 1.63564655[15] \\
~~~~ & ~~~~ & [(1,3)\{3,3,1\}]  & 
 1.68994048[20]  &   ~~~~      & ~~~~~ \\ 
\end{tabular}
\end{ruledtabular}
\end{table*}

We note that in several cases the nonspurious states are given by a single trial 
state as defined in Eq.\ (\ref{mol_trial_wf}). Indeed for $L=9$ the yrast state 
is a pure REM state, i.e., $\Phi^{\text{REM}}_{9}(1,3)$. For $L=11$ the single 
nonspurious state is the first excited state in the full spectrum,  
coinciding with the molecular vibration $\Phi^{\text{REM}}_{9}(1,3)Q_2$. 

Of particular interest is the $L=18$ case; it corresponds to the celebrated
$\nu=1/3$ fractional filling, which is considered \cite{laug8399} as the 
prototype of quantum liquid states. However, in this case we found (see Table 
\ref{ene_ferm_np4}) that the exact nonspurious solutions are linear 
superpositions of seven molecular states involving dipole $(\lambda=2)$ 
and octupole $(\lambda=3)$ vibrations over both the (0,4) and (1,3) 
configurations. Focusing on the yrast state with $L=18$, we found that its 
largest component is the
pure $\Phi^{\text{REM}}_{18}(0,4)$ REM state with a 0.9294 overlap with the EXD
solution; the contibutions of the remaining six states are much smaller, but 
they bring the overlap to precisely unity. Unlike the $\nu=1/2$ case of bosons, 
we stress that the fermionic Jastrow-Laughlin functions at all $\nu$'s exhibit 
less-than-unity overlaps \cite{laug8399,jainbook}.

Of great interest also is the $L=30$ ($\nu=1/5$) case, which in the
composite-fermion picture was found to be susceptible to a competition 
\cite{chan06} between crystalline and liquid orders. However, we found that the 
exact nonspurious states for $L=30$ are actually linear superpositions of the 
following 19 $[=D^{\text{TI}}(L=30)]$ RVM functions: 
$\Phi^{\text{REM}}_{6+4k}(0,4)Q_2^{12-2k}$, with $k=0,1,2,3,4,5,6$; 
$\Phi^{\text{REM}}_{6+3k}(1,3)Q_2^{12-3k/2}$, with $k=2,4,6$;
$\Phi^{\text{REM}}_{6+4k}(0,4)Q_3^{8-4k/3}$, with $k=0,3$; and
$\Phi^{\text{REM}}_{6+3k}(1,3)Q_3^{8-k}$, with $k=2,3,4,5,6,7,8$.
Diagonalization of the Coulomb interaction in the above TI subspace yielded an 
energy 0.25084902 $e^2/(\kappa l_B)$ per electron for the yrast state; this 
value agrees again, within machine precision, with the EXD result. 
The most sophisticated variants of the composite-fermion theory (including 
composite-fermion diagonalization (CFD), 
composite-fermion crystal (CFC), and mixed liquid-CFC states \cite{chan06,jeon07,
jainbook}) fall short in this respect. Indeed the following higher energies were 
found \cite{chan06}: 0.250863(6) (CFD), 0.25094(4) (mixed), 0.25101(4) (CFC).
The CFD basis is not translationally invariant. Consequently, to achieve 
machine-precision accuracy, the CFD will have to be performed
in the larger space of dimension $D^{\text{EXD}}(L=30)=169$.

{\it Conclusions.\/} $-$ 
The many-body Hilbert space corresponding to the translationally 
invariant part of the LLL spectra of small systems (whether fermions or bosons,
and for both low and high angular momenta) is spanned by the 
rovibrational molecular trial functions introduced in Eq.\ (\ref{mol_trial_wf}). 
The yrast and excited states for both short- and long-range
interactions can always be expressed as linear superpositions of these trial
functions. Thus the nature of strong correlations in the lowest Landau 
level reflects the spontaneous emergence of intrinsic point-group symmetries 
associated with rotations and vibrations of molecules of localized particles
arranged in concentric polygonal-ring configurations. We stress, particularly,
the validity and numerical superiority of the present molecular theory 
for {\it low\/} angular momenta, where fundamentally different 
"quantum-liquid" physical pictures (based on Laughlin, composite-fermion, and 
Pfaffian functions) have been assumed \cite{laug8399, jainbook, 
coop08} to apply.


\begin{thebibliography}{999}
\bibitem{note3}
An yrast state has the lowest enegy at a given angular momentum.
\bibitem{note2}
Use of an algebraic language like MATHEMATICA is necessary.
\end{thebibliography}
\end{document}